\documentclass{article}

\usepackage{graphicx} 
\usepackage{authblk}

\usepackage{booktabs}
\usepackage{color}
\usepackage[left=3 cm, right=3 cm, top=2.cm]{geometry}
\usepackage{amsmath}
\usepackage{changepage} 
\usepackage[numbers]{natbib}  

\usepackage{listings}
\usepackage{xcolor}
\usepackage{float}
\usepackage{parskip} 
\usepackage{multicol}

\usepackage{url}
\usepackage{hyperref}
\usepackage{lineno}
\usepackage{microtype}

\usepackage{natbib}
\usepackage{bm}
\usepackage{mathrsfs}
\usepackage{amssymb}
\usepackage{color}
\usepackage{amsmath}
\usepackage{tablefootnote}
\usepackage{tabularx}
\usepackage{caption}
\usepackage{multirow}
\usepackage{booktabs}
\usepackage{makecell}
\usepackage{caption}
\usepackage{float}

\usepackage{epstopdf}
\usepackage{tikz}

\title{Imaging Genetics Analysis of Alzheimer’s Disease}

\author[1]{Riddhik Basu }

\author[2]{Arkaprava Roy}

\author[3]{For The Alzheimer's Disease Neuroimaging Initiative  \thanks{ Data used in the preparation of this article were obtained from the Alzheimer's Disease Neuroimaging Initiative (ADNI) database (adni.loni.usc.edu). As such, the investigators within the ADNI contributed to the design and implementation of ADNI and/or provided data but did not participate in the analysis or writing of this report. A complete listing of ADNI investigators can be found at:
\url{http://adni.loni.usc.edu/wp-content/uploads/how_to_apply/ADNI_Acknowledgement_List.pdf} **Corresponding authors. Email: rbasu2@ncsu.edu or arkaprava.roy@ufl.edu. }   }

\affil[1]{Department of Statistics, North Carolina State University, Raleigh, NC, USA}
\affil[2]{Department of Biostatistics, University of Florida, Gainesville, FL, USA}

\date{}

\begin{document}
\maketitle

\vspace{2.5 cm}

\begin{abstract} 

Alzheimer’s disease (AD) is a progressive neurodegenerative disorder characterized by cognitive decline, structural brain changes, and genetic predispositions. This study leverages machine learning and statistical techniques to investigate the mechanistic relationships between cognitive function, genetic markers, and neuroimaging biomarkers in AD progression. Using data from the Alzheimer’s Disease Neuroimaging Initiative (ADNI), we perform both low-dimensional and high-dimensional analyses to identify key predictors of disease states, including cognitively normal (CN), mild cognitive impairment (MCI), and AD. Our low-dimensional approach utilizes multiple linear and ordinal logistic regression to examine the influence of cognitive scores, cerebrospinal fluid (CSF) biomarkers, and demographic factors on disease classification. The results highlight significant associations between Mini-Mental State Examination (MMSE), Clinical Dementia Rating Sum of Boxes (CDRSB), and phosphorylated tau levels in predicting cognitive decline. The high-dimensional analysis employs Sure Independence Screening (SIS) and LASSO regression to reduce dimensionality and identify genetic markers correlated with cognitive impairment and white matter integrity. Genes such as CLIC1, NAB2, and TGFBR1 emerge as significant predictors across multiple analyses, linking genetic expression to neurodegeneration. Additionally, imaging genetic analysis reveals shared genetic influences across brain hemispheres and the corpus callosum, suggesting distinct genetic contributions to white matter degradation. These findings enhance our understanding of AD pathology by integrating cognitive, genetic, and imaging data. Future research should explore longitudinal analyses and potential gene-environment interactions to further elucidate the biological mechanisms underlying AD progression.

\end{abstract}

\newpage

\section*{1. Introduction}
\addcontentsline{toc}{section}{Introduction}

Alzheimer's disease (AD) is a progressive neurodegenerative disorder that impairs memory, thinking, and behavior, leading to severe cognitive decline over time. It is the most common cause of dementia, accounting for 60–80\% of all dementia cases \cite{irwin2018healthy}. AD typically begins with mild memory loss and gradually progresses to more severe impairments, ultimately impacting a person’s ability to perform everyday tasks and maintain independence. The key pathological hallmarks of AD include the accumulation of amyloid-beta plaques and tau tangles in the brain, leading to neuronal damage and cognitive decline \cite{deture2019neuropathological}. Currently, there is no known cure, but early diagnosis and interventions can help manage symptoms and improve the quality of life for affected individuals.

Cognitive impairment presents itself in several stages, beginning with Cognitively Normal (CN), where individuals do not display any significant cognitive issues. This is followed by Mild Cognitive Impairment (MCI), a transitional stage between normal aging and Alzheimer's disease. MCI can be further categorized into early (EMCI) and late (LMCI) stages, based on the degree of impairment \cite{aisen2015alzheimers}. In its advanced form, cognitive decline progresses to AD, where patients experience significant memory loss, impaired reasoning, and behavioral disturbances. Neuropsychological exams are essential tools for assessing cognitive function and directly quantifying cognitive in diagnosed patients. These exams evaluate various domains such as memory, attention, executive function, and language, providing valuable insights for distinguishing normal aging, and differences between MCI versus AD \cite{nelson2008mci}. Examples of these exams include the Alzheimer’s Disease Assessment Scores (ADAS), the Mini-Mental State Exam (MMSE), and more \cite{mohs1988adas, folstein1975mmse}. As a commonly used assessment, the MMSE is scored on a 30-point scale, with higher scores indicating better cognitive function. Scores above 27 suggest normal cognition, scores between 24–27 indicate possible MCI, and scores below 24 typically indicate more severe impairments, such as in AD \cite{arevalo2015mmse}.

Dementia is associated with notable structural changes in the brain, including atrophy of the hippocampus and cortical regions responsible for memory and cognition \cite{wiseman2004hippocampal}. White matter integrity is often disrupted, and diffusion tensor imaging (DTI) studies have identified changes in Fractional Anisotropy (FA) \cite{kantarci2014white, mayo2019white}, a measure of white matter microstructure. Lower FA values reflect impaired neuronal connectivity, which is commonly observed in patients with AD and MCI \cite{tae2018diffusion}. These structural abnormalities correspond to cognitive decline, making FA an important biomarker for studying disease progression. Genetic research plays a crucial role in understanding the development and progression of Alzheimer's disease. Variants in genes such as apolipoprotein-E (APOE) allele \cite{troutwine2022apoe} 
have been strongly associated with increased risk for AD, as shown in Figure~\ref{fig:APOE4_Jitter} \textit{(Appendix A)}. Beyond APOE, genome-wide association studies (GWAS) have identified other genes involved in amyloid processing, neuroinflammation, and lipid metabolism that may contribute to AD pathology \cite{adgc2019genetic}. Exploring these genetic markers helps identify individuals at higher risk, informs personalized treatments, and advances the understanding of disease mechanisms.

In this paper, we conduct multiple analyses to better understand the relationship between cognitive function, genetic markers, and structural brain changes. First, we apply statistical techniques to assess the influence of key biomarkers on disease states (CN, MCI, AD.) Next, we perform an association study between genetic markers and disease states to identify relevant genetic factors. We then investigate the relationship between FA values and genetic markers to explore how brain structure is influenced by genetic predisposition. Finally, we analyze common genes across disease states to identify shared pathways contributing to cognitive decline. In Section 2, we describe the methodology, including data collection, and preprocessing steps, followed by Section 3, introducing our statistical methods to be used in the analysis. Section 4 discusses the findings of our overall genetic and imaging analyses, as well as their implications. Finally, in Section 5, we conclude with a summary of the key results including potential limitations and provide directions for future research.

\section*{2. ADNI-Data}
\addcontentsline{toc}{section}{ADNI-Data}

The Alzheimer’s Disease Neuroimaging Initiative (ADNI) is an ongoing longitudinal cohort study designed to develop clinical, imaging, and genetic markers for early detection and tracking of AD \cite{petersen2010adni}. Eligible participants, aged 55-90 and in generally good health, were enrolled with either memory concerns or normal cognition. Each participant underwent comprehensive assessments, including cognitive testing, imaging, genetic evaluations, and both invasive and non-invasive medical procedures. Follow-up visits were conducted approximately every six months. The TADPOLE Challenge provided original data conducted from ADNI.

\subsection*{2.1 Cohort Description and Pre-Processing}
\addcontentsline{toc}{subsection}{Cohort Description and Pre-Processing}

Our study cohort consists of baseline visits from three phases: ADNI-1, ADNIGO, and ADNI-2 \cite{petersen2010adni, weber2021adnigo, beckett2015adni2}. All patients included in the analysis were de-identified and exhibited varying degrees of cognitive impairment. Although ADNI provides follow-up data, we focused exclusively on baseline data to better understand the population-level characteristics of cognitively impaired subjects. For data preprocessing, visualization, and statistical analysis, we utilized a combination of tools. Preprocessing tasks were conducted using Python 3, employing packages such as Numpy \cite{harris2020numpy}, Pandas \cite{mckinney2010pandas}, and Matplotlib \cite{hunter2007matplotlib} for data manipulation. For statistical analysis, we used R statistical software version 4.3.3. The primary goal of this study is to identify biomarkers associated with the acceleration of AD progression. After selecting baseline-only patients and removing incomplete or censored data, our final cohort consisted of 1,631 patients (819-ADNI1, 129-ADNIGO, and 683-ADNI2). Certain biological factors were excluded from analysis due to incomplete data or patient refusal. The selected predictors included are listed below in Table \ref{tab:variables}. A more comprehensive summary of biomarker statistics is provided in \textit{Appendix A} Table~\ref{tab:Biomarkers_Tab}.

\begin{table}[H]
    \centering
    \renewcommand{\arraystretch}{1.2} 
    \begin{tabular}{ l p{9.5 cm} }
        \toprule
        \textbf{Category} & \textbf{Description} \\
        \midrule
        Risk Factors & Age, Years of education, and APOE-4 genetic stats\\
        Cognitive Exams & ADAS-Cog (ADAS-11 and ADAS-13), Clinical Dementia Rating Sum of Boxes (CDRSB), Mini Mental State Exam (MMSE), and Rey Auditory Verbal Learning Test \\
        CSF Measures & Amyloid-beta, tau, and phosphorylated tau levels \\
        Genetic Expression & Locus-Links, Probe Sets, Genes \\
        Diffusion Tensor Imaging & Fractional Anisotropy (FA), White Matter Integrity\\
        \bottomrule
    \end{tabular}
    \caption{Summary of Key Measures and Biomarkers}
    \label{tab:variables}
\end{table}

Our secondary objective is to investigate genetic associations with disease progression. We analyze gene expression data in relation to the biomarkers identified in our primary analysis and explored correlations between gene expression and brain structure changes by examining Fractional Anisotropic levels in diffusion tensor imaging (DTI) data. This analysis is insightful in identifying clusters of white matter associated with rapid neurodegeneration \cite{poulakis2021white}.

The gene expression data include probe sets, Locus Link (Gene-ID), with a combined total of 49,386 genes. We filtered and merged the gene expression records with the rest of the data, yielding a sample of 468 patients with gene expression data and a sub-sample of 104 patients with both DTI imaging and genetic data.

\section*{3. Statistical Methods}
\addcontentsline{toc}{section}{Statistical Methods}

We now review some of the low- and high-dimensional statistical methods that are considered in our analysis.


\subsection*{3.1 Low-Dimensional Analysis}
\addcontentsline{toc}{subsection}{Low-Dimensional Analysis}

\subsubsection*{3.1.1 Multiple Linear Regression}
\addcontentsline{toc}{subsubsection}{Multiple Linear Regression}

A multiple linear regression fits the following model for an outcome $Y$ and an array of $p-1$ predictors $X_1, X_2,...,X_{p-1},$ assuming that the expectation of the error is 0, $E(e_i) =0$

\begin{equation}
    Y_i = \beta_0 + \beta_1X_{i,1} + \beta_2X_{i,2} + \dots + \beta_{p-1}X_{i,p-1}+e_{i}
    \label{eq:Linear_regression}
\end{equation}

For the purposes of this analysis, we prioritize identifying and retaining predictors that exhibit statistically significant associations with the outcome, ensuring the model's accuracy and interpretability while reducing the risk of overfitting. This careful selection of predictors is crucial for drawing valid inferences from the regression model. Later, we will augment this by implementing a penalty to shrink lesser-degree variables.

\subsubsection*{3.1.2 Ordinal Logistic Regression}
\addcontentsline{toc}{subsubsection}{Ordinal Logistic Regression}

The ordinal model is a statistical technique used to model the relationship between an ordinal dependent variable and independent variables. It utilizes logistic regression (like a link function) to measure the likelihood will fall into one of the categories. Our key assumption is that the relationship between predictors and log-odds of outcomes is proportional across all thresholds (proportional odds assumption.) \cite{yee2010ordinal}

\begin{equation}
    log\bigg[ \frac{P(Y \leq j)}{P(Y>j)}  \bigg] = logit\bigg[ P(Y \leq j)  \bigg]
    \label{eq:Ordinal_regression}
\end{equation}

$$
logit\big( P(Y \leq j) \big) = \beta_0 - n_1x_1-n_2x_2 - ... - n_{p-1}x_{p-1}
$$

Where: The function holds $Y$ outcomes, $j$ categories and $p-1$ predictors.

In our case, we can probabilistically assess how biomarkers, such as cognitive exams or genetic data, influence disease state progression. This allows us to understand and visualize how various biomarkers impact the likelihood of being in different stages of the disease, offering insights into the probabilistic nature of disease progression.

\subsection*{3.2 High-Dimensional Analysis}
\addcontentsline{toc}{subsection}{High-Dimensional Analysis}

\subsubsection*{3.2.1 Screening}
\addcontentsline{toc}{subsubsection}{Screening}

Due to the vast amount of genetic information, we reduce the dimensionality. To achieve this, we applied the Sure Independence Screening method \cite{fan2008sure}. 
SIS effectively selects a subset of the most important predictors by ranking them based on their marginal association with the outcome variable. By focusing on top-ranking features, SIS ensures that we do not lose important predictors, and we can then apply more refined variable selection techniques in a computationally efficient way.

We applied the Least Absolute Shrinkage and Selection Operator (LASSO) penalty-based regression methods. LASSO is particularly beneficial in our case because it automates model selection through its shrinkage effect, as well as preventing over-fitting \cite{tibshirani1996lasso}. It effectively reduces the coefficients of less significant genes to zero, concentrating on relevant, non-zero predictors.

We discuss the two models, considered in this paper.  The first is the ``multi-task learning'' model, and the second is for multi-category outcomes.

\subsubsection*{3.2.2 Multivariate Response}
\addcontentsline{toc}{subsubsection}{Multivariate Response}

The multi-response Gaussian family is particularly useful when there are several (correlated) responses, also known as the “multi-task learning” problem \cite{hastie2024glmnet}. In this case, a covariate is either included in the model for all the responses or excluded for all the responses. 

\begin{equation}
    \begin{pmatrix}
    Y_{1,1} & \cdots & Y_{1,m} \\
    \vdots & \ddots & \vdots \\
    Y_{n,1} & \cdots & Y_{n,m}
    \end{pmatrix}
    =
    \begin{bmatrix}
    x_{1,1} & \cdots & x_{1,p} \\
    \vdots & \ddots & \vdots \\
    x_{n,1} & \cdots & x_{n,p}
    \end{bmatrix}
    \begin{bmatrix}
    \beta_{1,1} & \cdots & \beta_{1,m} \\
    \vdots & \ddots & \vdots \\
    \beta_{p,1} & \cdots & \beta_{p,m}
    \end{bmatrix}
    +
    \begin{pmatrix}
    \varepsilon_{1,1} & \cdots & \varepsilon_{1,m} \\
    \vdots & \ddots & \vdots \\
    \varepsilon_{n,1} & \cdots & \varepsilon_{n,m}
    \end{pmatrix}
\end{equation}

\[
\rightarrow Y_{n \times m} = X_{n \times p} \, \beta_{p \times m} + \varepsilon_{n \times m}
\]


The loss function consists of two main terms. The first term is the squared Frobenius norm of the residuals, measuring the difference between the observed and predicted responses across multiple outcomes. The second term is the group LASSO penalty, encouraging sparsity by driving some coefficients of some groups to zero.

\begin{equation}
    \min_{(\beta_0, \beta) \in \mathbb{R}^{(p+1) \times K}} \hspace{.2 cm} \frac{1}{2N} \sum_{i=1}^{N} \| y_i - \beta_0 - \beta^T x_i \|_F^2 + \lambda \sum_{j=1}^{p} \|\beta_j\|_2
\end{equation}
where $\|\cdot\|_F$ and $\lambda$ stand for the Frobenius norm and group lasso penalty, respectively.



\subsubsection*{3.2.3 Multinomial Response}
\addcontentsline{toc}{subsubsection}{Multinomial Response}

The multinomial model is particularly useful for classification problems where the response variable can take multiple categorical outcomes rather than just binary labels \cite{hastie2024glmnet}. Unlike one-vs-all approaches, the multinomial model jointly models all class probabilities using the softmax function, ensuring that the probabilities sum to one. This allows the model to capture relationships between different categories rather than treating them independently. 

Suppose the response variable has $J$ levels, $G={1,2,…,J}$. Then our multinomial model is:

\begin{equation}
    P(G = j \mid X = x) = \frac{e^{\beta_{0,j} + \beta_j^T x}}{\sum_{\ell=1}^{J} e^{\beta_{0,\ell} + \beta_\ell^T x}}.
\end{equation}

Let $Y$ be the $N \times K$ indicator response matrix, with elements $y_{i\ell}=I(g_i = \ell)$. Then the elastic net penalized negative log-likelihood function becomes:

\begin{equation}
    \ell(\{\beta_{0k}, \beta_k\}_{k=1}^{K}) = -\left[ \frac{1}{N} \sum_{i=1}^{N} \left( \sum_{k=1}^{K} y_{i,k} \log(\beta_{0k} + x_i^T \beta_k) - \log\left( \sum_{\ell=1}^{K} e^{\beta_{0\ell} + x_i^T \beta_\ell} \right) \right) \right] + \lambda \sum_{j=1}^{p} \|\beta_j\|_1.
\end{equation}

The first term represents the negative log-likelihood for a multinomial classification model, which calculates the likelihood of observing the correct class labels, \( y_{i,k} \), given the predicted probabilities modeled using the softmax function. The sum inside the logarithm computes the probability for each class, and the negative log-likelihood penalizes incorrect classifications. The second term is the LASSO penalty (\(\ell_1\)-norm), which regularizes the model by enforcing sparsity in the coefficients. This term shrinks the coefficients,  selecting only the most important features.

\section*{4. Results}
\addcontentsline{toc}{section}{Results}

With three distinct analytical objectives, we begin with a full cohort of 1,631 patients, which we use for all low-dimensional analyses. From this cohort, we identify a subset of 468 patients with available genetic information. Among these, a further subset of 104 patients also has Fractional Anisotropy (FA) imaging data available. This sub-subset is used for imaging-informed kernel-based analyses. Across all levels of analysis, a p-value threshold of $< 0.05$ is considered statistically significant. We apply the following \texttt{R} packages for our analysis: \texttt{glmnet} \cite{glmnet1, glmnet2, glmnet3}, \texttt{SIS} \cite{saldana2018sis}, \texttt{corrplot} \cite{corrplot2021}, \texttt{ggplot2} \cite{RpackGGplot}, \texttt{MASS} \cite{RpackageMASS}, \texttt{car} \cite{RpackCar}, and \texttt{gtsummary} \cite{gtsummary}.

\subsection*{4.1 Low-dimensional Analysis}
\addcontentsline{toc}{subsection}{Low-dimensional Analysis}

\subsubsection*{4.1.1 Regressing Disease State on Cognitive Scores}
\addcontentsline{toc}{subsubsection}{Regressing DS on Cognitive Scores}

\begin{figure}[H]
    \centering
    \includegraphics[width = .8\linewidth, height = 8.50cm]{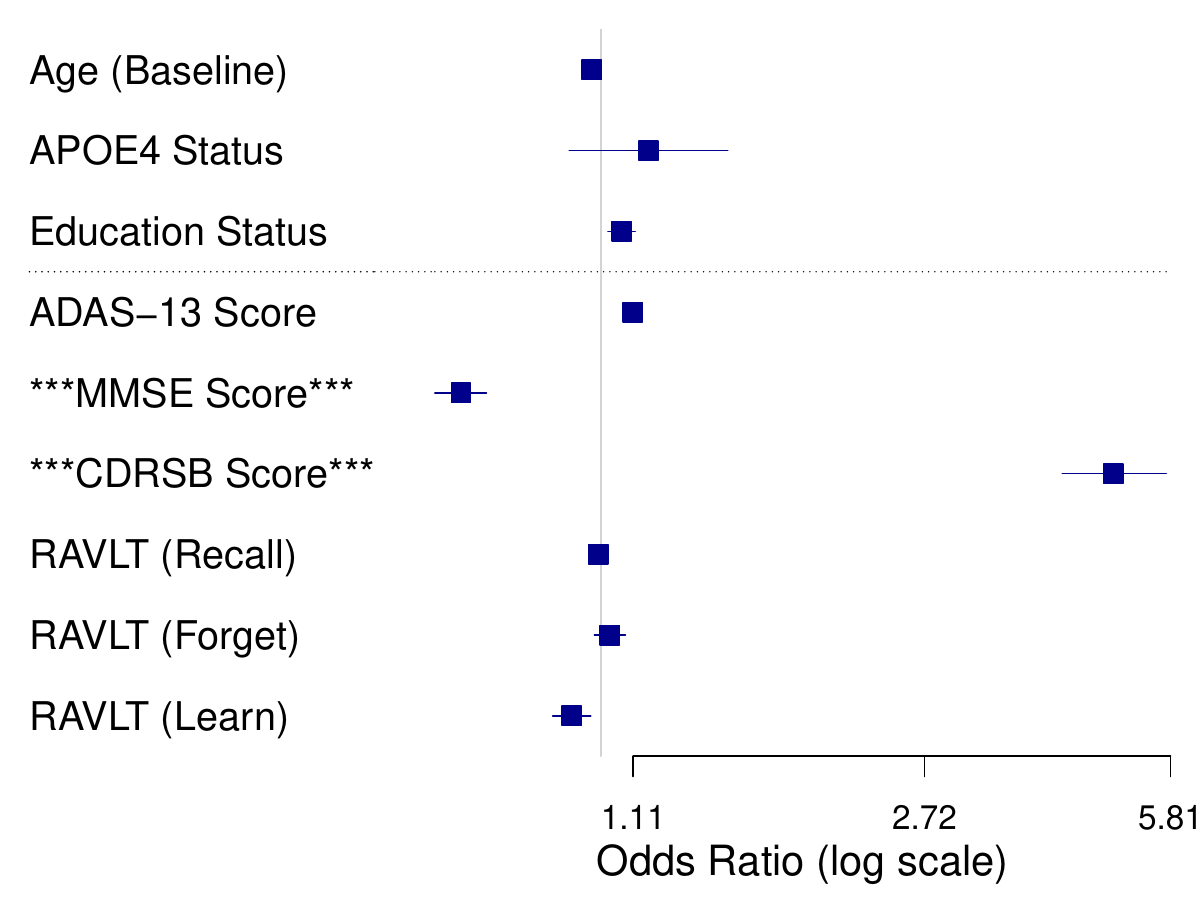}
    \caption{Estimated regression coefficients with 95\% confidence intervals from an ordinal regression model, where disease state (Normal → EMCI → LMCI → AD) is the outcome and the listed variables on the left are the covariates.}
    \label{fig:regoutput1}
\end{figure}

We regress disease states on a set of cognitive test scores and other risk factors using an ordinal regression model. Our selection of variables is based on known demographic factors as well as the most common assessments administered to dementia patients. Figure~\ref{fig:regoutput1} illustrates odds ratios in the log-scale from the ordinal regression model.  

The results highlight statistically significant associations between several cognitive exam scores and disease progression, as visualized in Figure \ref{fig:regoutput1}. Among the most influential predictors, the Mini-Mental State Examination (MMSE) and the Clinical Dementia Rating–Sum of Boxes (CDRSB) display the smallest and largest odds ratios, respectively. This finding is consistent with clinical expectations, as lower MMSE scores and higher CDRSB scores reflect worsening cognitive function and more advanced stages of Alzheimer’s disease \cite{andrews2019mcid, lanctot2024cdrmmse}. In contrast, APOE4 carrier status shows a positive and statistically significant association with disease progression \cite{qian2021apoegenotype}. Variables such as education level, baseline age, and RAVLT (Rey Auditory Verbal Learning Test) scores do not reach statistical significance (p-values $>$ 0.05) in the multivariate model. Notably, MMSE and CDRSB scores demonstrate extreme magnitudes, indicating that each per-unit increase in MMSE provides a protective effect, while increments in CDRSB signal worsening cognitive impairment. Further analysis is required to investigate these biomarkers.

\subsubsection*{4.1.2 Cognitive Scores Analysis}
\addcontentsline{toc}{subsubsection}{Cognitive Scores Analysis}

Neuropsychological exams provide a direct measure of cognitive decline, assessing various skills such as memory, language, and visual processing. These tests help clinicians gauge a patient’s overall cognitive awareness. In our analysis, we focus on the CDRSB and MMSE as response variables within a multivariable regression framework. We include disease state as an explanatory variable to explore changes across individual conditions. Due to multicollinearity, no other neuropsychological tests are incorporated into the model. We first analyze the CDRSB to identify significant factors, followed by a regression of the MMSE.

\subsection*{Multivariable Regression Analysis of CDRSB}
Table~\ref{tab:CDRSB} presents the multivariable linear regression estimates for regressing the Clinical Dementia Rating scale Sum of Boxes (CDRSB) on the combined CSF measurements and disease states.

\begin{table}[h]
    \centering
     \caption{Multivariable Linear Regression Estimates for CDRSB regressed on the biomarkers, listed in the first column with sample size N = 1,113}
    \label{tab:CDRSB}
    \begin{tabular}{lcccc}
        \toprule
        \textbf{Biomarker} & \textbf{Estimate} & \textbf{Std. Error} & \textbf{T-value} & \textbf{P-Value} \\
        \midrule
        \multicolumn{5}{c}{\textbf{Demographics}} \\
        Baseline Age (years) & -4.919 $\times 10^{-4}$ & 0.004 & -0.119 & 0.905 \\
        APOE-4 Genetic Status & 3.491 $\times 10^{-3}$ & 0.068 & 0.051 & 0.959 \\
        \midrule
        \multicolumn{5}{c}{\textbf{Disease State}} \\
        Early Mild CI & 2.909 & 0.072 & 40.512 & $< 2 \times 10^{-16}$ \\
        Late Mild CI & 0.764 & 0.061 & 12.601 & $< 2 \times 10^{-16}$ \\
        Alzheimer’s Disease & 0.768 & 0.058 & 13.363 & $< 2 \times 10^{-16}$ \\
        \midrule
        \multicolumn{5}{c}{\textbf{Cerebrospinal Fluid (CSF)}} \\
        Phosphorylated Tau levels & 5.108 $\times 10^{-3}$ & 0.002 & 2.265 & 0.024 \\
        Amyloid-Beta levels & -1.514 $\times 10^{-4}$ & 5.792 $\times 10^{-5}$ & -2.614 & 0.009 \\
        \bottomrule
    \end{tabular}
\end{table}

The multivariable linear regression model assessing Clinical Dementia Rating Sum of Boxes (CDRSB) yielded an Adjusted \( R^2 \) of 0.699, indicating that approximately 70\% of the variability in CDRSB is explained by the model. Central to this analysis were the cerebrospinal fluid (CSF) biomarkers, with Phosphorylated Tau levels (PTAU) and Amyloid-Beta levels (ABETA). Since the correlation between PTAU and TAU levels was around 98\%, we only included PTAU. 

\subsection*{Multivariable Regression Analysis of MMSE}
Table~\ref{tab:MMSE} presents the multivariable linear regression estimates for regressing the Mini-Mental State Examination (MMSE) on the combined CSF measurements and disease states.

\begin{table}[h]
    \centering
       \caption{Multivariable Linear Regression Estimates for MMSE regressed on the biomarkers, listed in the first column with sample size N = 1,113}
    \label{tab:MMSE}
    \begin{tabular}{lcccc}
        \toprule
        \textbf{Biomarker} & \textbf{Estimate} & \textbf{Std. Error} & \textbf{T-value} & \textbf{P-Value} \\
        \midrule
        \multicolumn{5}{c}{\textbf{Demographics}} \\
        Baseline Age (years) & -0.029 & 6.842 $\times 10^{-3}$ & -4.232 & $2.51 \times 10^{-5}$ \\
        APOE-4 Genetic Status & -0.072 & 0.113 & -0.640 & 0.522 \\
        \midrule
        \multicolumn{5}{c}{\textbf{Disease State}} \\
        Early Mild CI & -3.780 & 0.119 & -31.843 & $< 2 \times 10^{-16}$ \\
        Late Mild CI & -1.500 & 0.100 & -14.960 & $< 2 \times 10^{-16}$ \\
        Alzheimer’s Disease & -0.594 & 0.095 & -6.245 & $6.04 \times 10^{-10}$ \\
        \midrule
        \multicolumn{5}{c}{\textbf{Cerebrospinal Fluid (CSF)}} \\
        Phosphorylated Tau levels & -0.012 & 0.004 & -3.185 & 0.001 \\
        Amyloid-Beta levels & 2.669 $\times 10^{-4}$ & 9.575 $\times 10^{-5}$ & 2.788 & 0.005 \\
        \bottomrule
    \end{tabular}
 
\end{table}

Similarly, the multivariable linear regression model for MMSE yielded an Adjusted \( R^2 \) of 0.618, indicating that 62\% of the variability in MMSE is explained by the model. CSF biomarkers were again key, with PTAU showing a significant negative association with cognitive performance, highlighting its role in cognitive decline. Conversely, Amyloid-Beta levels demonstrated a modest positive association with MMSE scores, consistent with their role in mitigating amyloid-related pathology.

\subsection*{4.2 High-Dimensional Analysis}
\addcontentsline{toc}{subsection}{HD Analysis}

\subsubsection*{4.2.1 Regressing CS on Gene Expression}

\addcontentsline{toc}{subsubsection}{Regressing CS on Gene Expression}

Recognizing the importance of MMSE and CDRSB as key markers of cognitive decline, we examine their associations from a genetic perspective. To address the high dimensionality of the genetic data, we apply a marginal screening technique using the Sure Independence Screening method \cite{Mishra2020AIAD}, which ranks all predictors and retains only the top variables for further analysis. We use our genetic subsample (n = 468), we perform SIS with the default setting of 10 partitions (nfolds = 10). We then fit a linear model with a Gaussian family, regressing the selected cognitive biomarkers (MMSE and CDRSB) on the screened genetic features. Only genes with non-zero coefficients in this final model are considered as potential predictors of cognitive decline.

\begin{figure}[H]
    \centering    
    \includegraphics[width = .95\linewidth, height = 7.25cm]{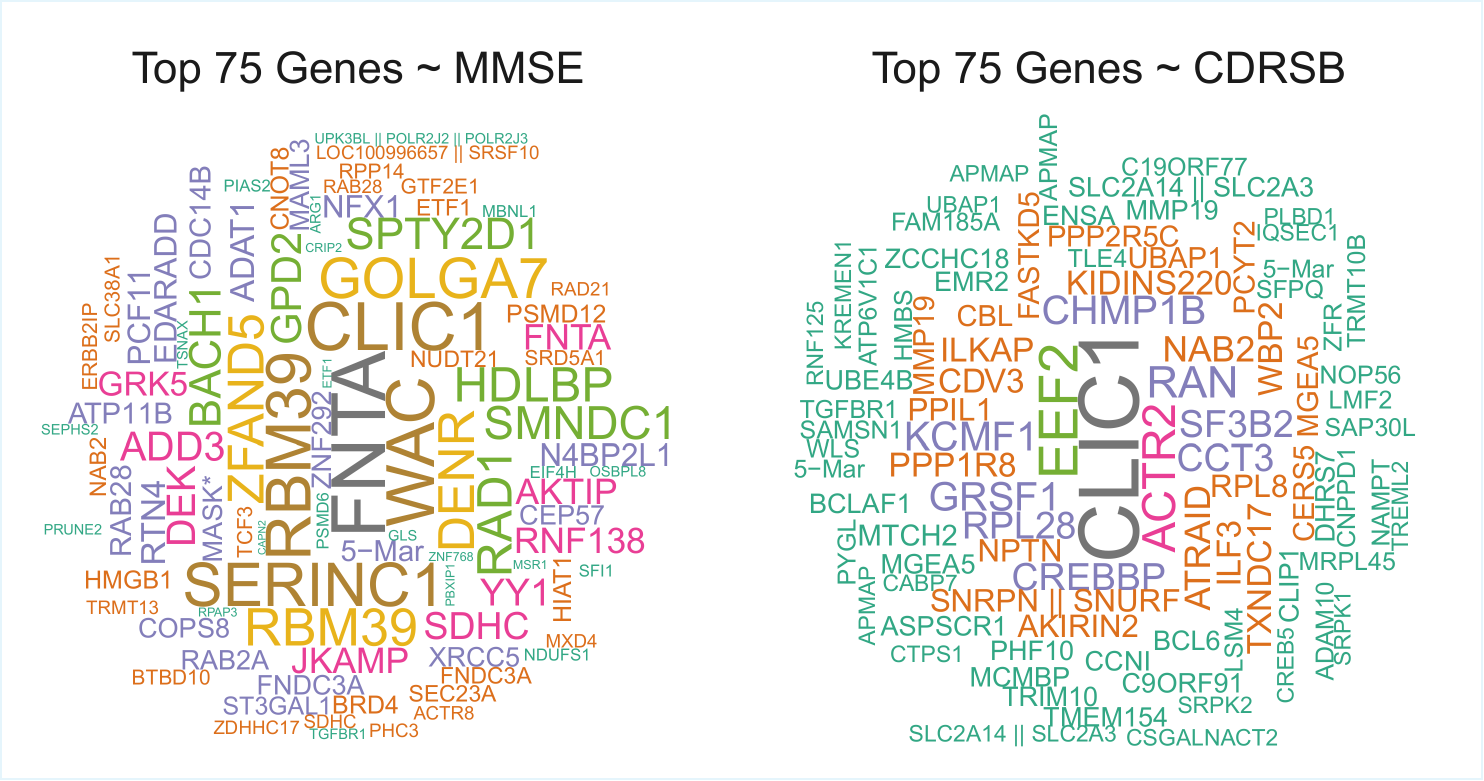} \caption{Word clouds of the top genes associated with cognitive decline, selected using a two-step procedure for both MMSE and CDRSB scores. First, Sure Independence Screening (SIS) reduced the high-dimensional feature space, and then LASSO regularization in Gaussian regression models with cross-validated lambda identified the most relevant genes. Only genes with non-zero coefficients in the final models were included, highlighting the strongest predictors of cognitive outcomes.} 
    \label{fig:Top75} 
\end{figure}

Figure \ref{fig:Top75} illustrates the top 75 genes identified based on the estimated regression coefficients, with the size of each gene name proportional to the magnitude of the coefficients in relation to cognitive function. For MMSE and CDRSB as responses to capture cognitive decline across disease stages, four key genes—\textit{NAB2}, \textit{5-MAR}, \textit{TGFBR1} \cite{Bosco2013}, and \textit{CLIC1}—emerge as statistically significant in both models, with CLIC1 \cite{carlini2020clic1} showing the greatest effect.

\subsubsection*{4.2.2 Regressing Disease State on Gene Expression}
\addcontentsline{toc}{subsubsection}{Regressing Disease State on Gene Expression}

We repeated the analysis to identify key genes across disease progression stages. Since SIS does not support multinomial models, we apply SIS, treating each pairwise comparison between disease stages (CN, EMCI, LMCI, AD) as separate binary classifications \cite{mahmud2023detection, sener2024categorization}. After SIS screening, we applied cross-validation using a multinomial model to capture genetic effects varying by disease stage. This approach allowed detection of genes significant in some transitions but not others. Finally, we compiled union and intersection sets of genes with non-zero coefficients across comparisons.

\begin{figure}[H]
    \centering
    \includegraphics[width = .80\linewidth, height = 9.50cm]{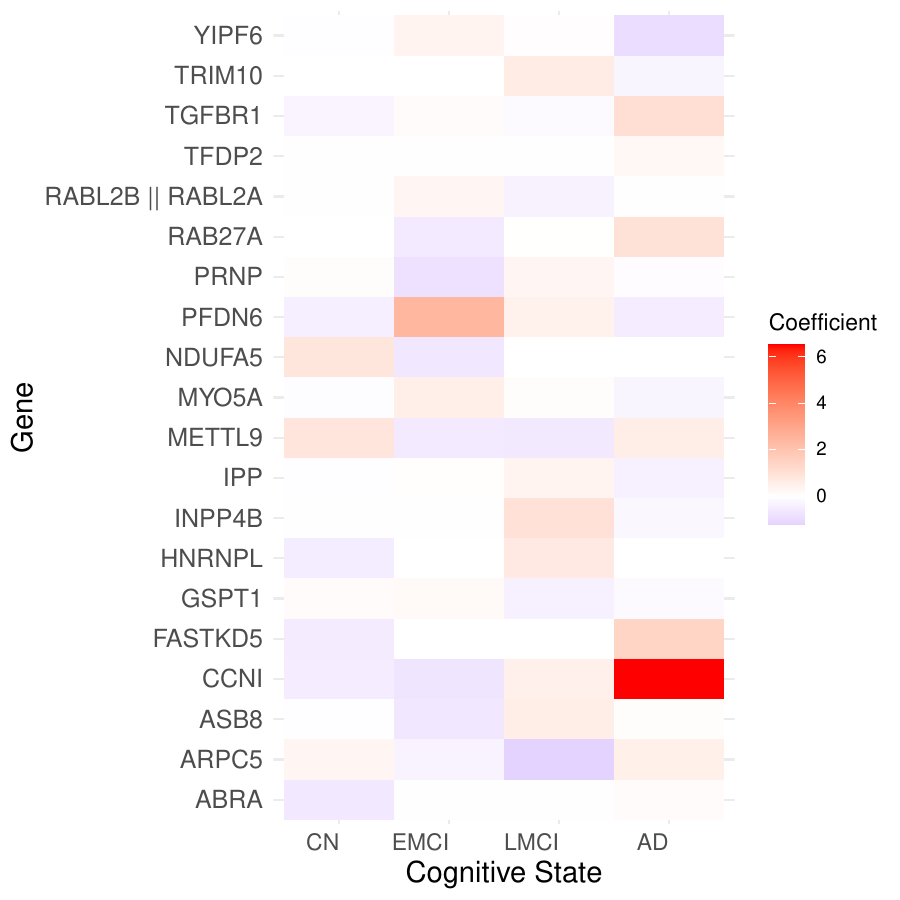}
    \caption{Estimated regression coefficients for genes with non-zero effects from LASSO-regularized multinomial logistic regression of disease state on gene expressions. Listed genes are the result of the intersection between each of the 6 pairwise groups between the 4 disease states.}
    \label{fig:DX_blPIC}
\end{figure}

Figure \ref{fig:DX_blPIC} highlights the associations between intersecting genes and cognitive states, with color gradients indicating effect magnitude. Notably, CCNI \cite{Hirabayashi2025} shows a strong positive association with Alzheimer's Disease, while PRNP has moderate associations in early to late cognitive decline stages. In contrast, genes like YIPF6 and RAB27A show negligible associations. These findings suggest that a gene which shows high effect at earlier or later stages of impairment may display a null effect at different stages. As such, finding intersections of the predictors reinforces our notion that a significant effect occurs in underlying mechanisms \cite{Zakeri2020}. In relation to the cognitive exam analyses from the previous step, we note that the following genes—TGFBR1, XRCC5, and CCNI—are correlated with both disease progression and individual cognitive scorings. It is important to note that genetic effects may not reflect a constant effect as disease state progresses.

\subsubsection*{4.2.3 Imaging Genetic Analysis}
\addcontentsline{toc}{subsubsection}{Imaging Genetic Analysis}

For our image analysis, we categorize the brain sections into three groups: the Corpus Callosum (CC), Right Brain, and Left Brain. The CC connects the two cerebral hemispheres and is crucial for higher cognitive functions. It is often one of the first regions to show signs of neurodegeneration. The left hemisphere handles language processing, analytical thinking, and logic, while the right hemisphere is more involved in spatial awareness and emotional interpretation. Damage to these structures can lead to disruptions in communication between the hemispheres, progressing neurodegenerative mechanisms \cite{dorion2022brain, zhang2023corpus}.

\begin{figure}[H]
    \centering

    \includegraphics[width=0.9\linewidth, height=7.75cm]{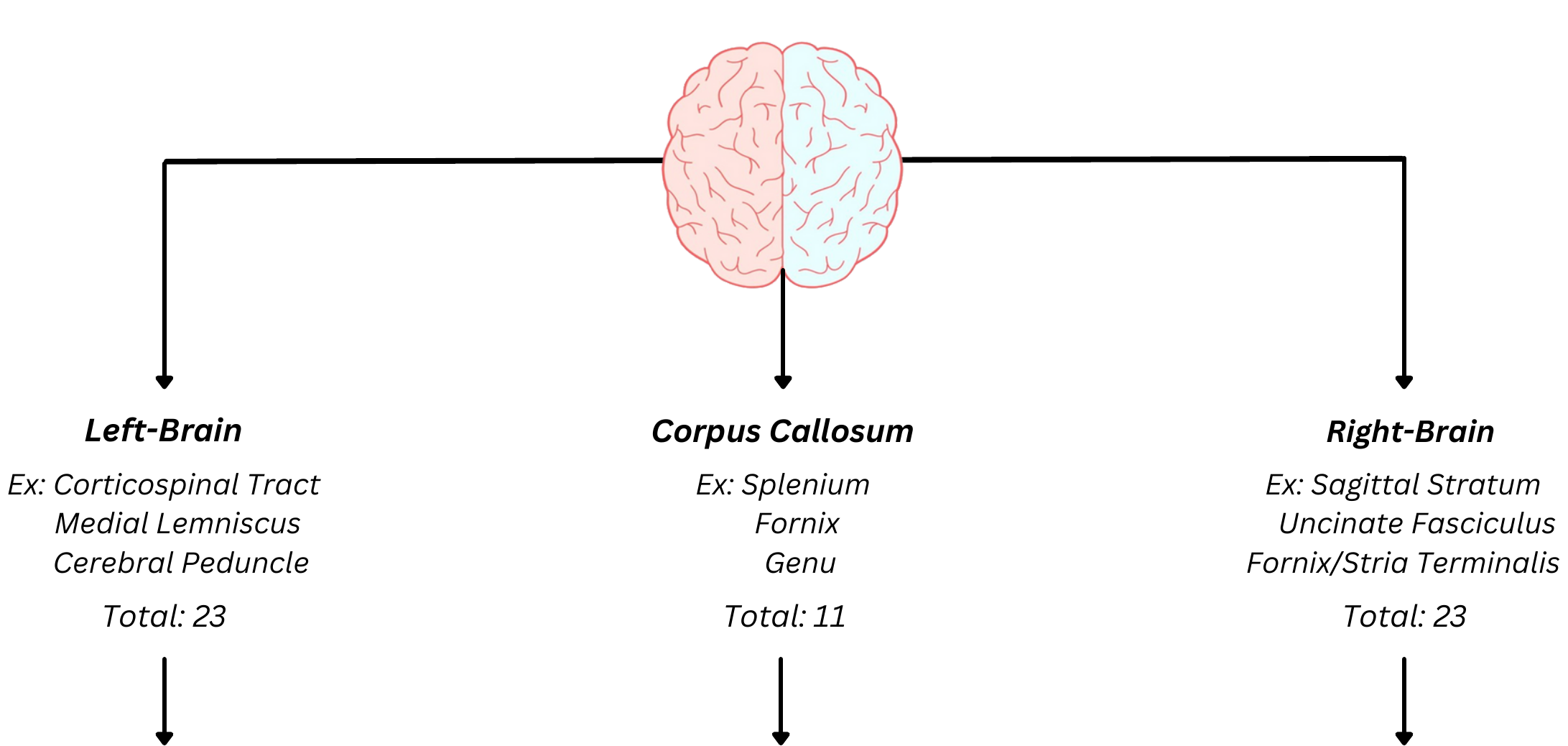}

    \vspace{-0.75 cm}  

    \[
    \underbrace{
    \begin{bmatrix}
        Y_{1,1}^{\text{LB}} & \cdots & Y_{1,23}^{\text{LB}} \\
        \vdots & \ddots & \vdots \\
        Y_{104,1}^{\text{LB}} & \cdots & Y_{104,23}^{\text{LB}}
    \end{bmatrix}}_{\text{Left Hemisphere}}
    \hspace{1.95cm}  
    \underbrace{
    \begin{bmatrix}
        Y_{1,1}^{\text{CC}} & \cdots & Y_{1,11}^{\text{CC}} \\
        \vdots & \ddots & \vdots \\
        Y_{104,1}^{\text{CC}} & \cdots & Y_{104,11}^{\text{CC}}
    \end{bmatrix}}_{\text{Corpus Callosum}}
    \hspace{2.05cm}  
    \underbrace{
    \begin{bmatrix}
        Y_{1,1}^{\text{RB}} & \cdots & Y_{1,23}^{\text{RB}} \\
        \vdots & \ddots & \vdots \\
        Y_{104,1}^{\text{RB}} & \cdots & Y_{104,23}^{\text{RB}}
    \end{bmatrix}}_{\text{Right Hemisphere}}
    \]

    \caption{Schematic diagram of the region-specific outcomes, with 23 subregions in the left and right hemispheres of the brain and 11 subregions in the corpus callosum. The matrices below show example data for each region.}
    \label{fig:Brain_Schematic}
\end{figure}

We have 57 FA image responses, with values ranging from 0 to 1. For fitting genetic predictors to these image responses, we apply SIS and the LASSO penalty. To account for heavy collinearity, we first apply a logit transformation, and then split the responses into their respective brain categories and fit them using multi-Gaussian. Due to the relatively small size of our subsample (n = 104), we adjust our cross-validation to 4 folds (nfolds = 4). After the selection process, we fit the significant genes across the 57 responses into Multivariate Gaussian models for each brain section. This results in three matrices, each corresponding to a set of non-zero gene predictors associated with a specific brain section, show in Figure \ref{fig:Brain_Schematic}.

In our final analysis, we set a threshold (e.g., the top 50 genes) for each brain section and identify predictors that intersect across multiple categories. This step involves examining the intersections with genetic predictors deemed significant from the initial genetic analysis. We observe that many genetic indicators overlap across the brain sections. For the Left and Right hemispheres, approximately 200 genes are associated with both, whereas about 80 genes are strongly associated with both hemispheres and the Corpus Callosum.

\begin{figure}[H]
    \centering
    \includegraphics[width = 1.0\linewidth, height = 6.35cm]{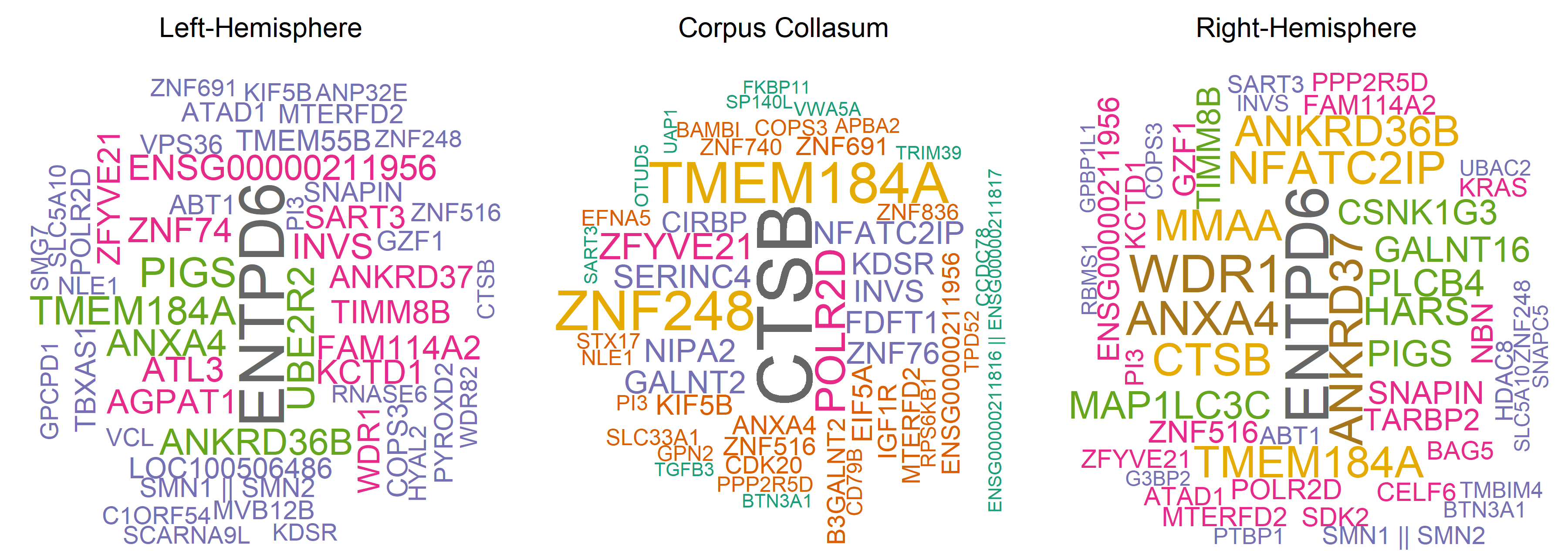}
    \caption{Word cloud visualizations of the top genes associated with different brain regions (Corpus Callosum, left and right hemispheres) identified using a multi-response LASSO regression with cross-validated lambda selection. Genes were selected based on non-zero coefficients from the regularized model across multiple responses, and the word cloud reflects both the union and intersection of genes with the strongest overall contributions to the brain region responses.}
    \label{fig:3_BrainSections}
\end{figure}

Figure \ref{fig:3_BrainSections} indicates the top 50 genes selected from the multi-Gaussian fitting process, with the size of each gene name reflecting its significance in relation to white matter integrity. Using the FA measurements to capture genetic effect, many predictors emerge as statistically significant in all three brain regions. Some of the listed genes include CTSB \cite{Hook2023CTSB}, TMEM184A, POLR2D \cite{Chum2024POLR}, INVS, ANXA4 \cite{White2024ANXA}, and others.

Our results suggest that the top genes in the two hemispheres share more commonalities compared to those in the corpus callosum. For instance, ENTPD6 is identified as significant in measuring white matter in the hemispheres but shows no significance in the CC. This finding may indicate that the genetic mechanisms governing white matter in the hemispheres are more uniform or shared, reflecting their complementary roles in processing information and supporting localized brain functions \cite{Zhao2021WhiteMatter}. In contrast, the gene TMEM184A demonstrates significance across all three regions, though with weaker effect sizes. Genes like TMEM184A, which are consistently significant across multiple brain regions, can serve as stronger indicators of white matter degradation and may hold greater potential as predictive markers.

\section*{5. Concluding Remarks and Future Directions}
\addcontentsline{toc}{section}{Concluding Remarks and Future Directions}

Our analysis provides an integrative view of Alzheimer’s disease progression by combining clinical, genetic, and imaging markers within both low- and high-dimensional statistical frameworks. Among the cognitive measures, lower MMSE scores and higher CDRSB scores consistently tracked with more advanced disease states, in line with clinical expectations and prior studies. Importantly, APOE4 carrier status emerged as a significant predictor of disease progression, reinforcing its established role as a genetic risk factor. In the high-dimensional setting, feature selection highlighted several candidate genes (e.g., CTSB, ANXA4, TMEM184A) that were associated with measures of white matter integrity across multiple brain regions. While individual genes such as ENTPD6 demonstrated hemisphere-specific effects, other predictors appeared consistently across regions, suggesting both localized and shared mechanisms of neurodegeneration. This hemispheric asymmetry aligns with prior evidence that the left and right hemispheres differentially contribute to language, spatial reasoning, and emotion, and that disruption of white matter pathways can compromise interhemispheric communication.

Taken together, these findings suggest that the mechanisms driving Alzheimer’s disease progression may not be uniform across the brain. Genes with consistent associations across multiple regions (e.g., TMEM184A) may serve as broad markers of white matter degeneration, whereas region-specific genes could reflect localized vulnerabilities. Such heterogeneity highlights the value of statistical frameworks that can capture both shared and region-dependent effects. A key strength of our work lies in the methodological integration of regression models, sure independence screening, and penalized regression for high-dimensional data. This approach allowed us to identify robust predictors from diverse data modalities while controlling for model complexity. By reframing Alzheimer’s research as a problem of variable selection across heterogeneous data types, our study contributes a generalizable framework that could be extended to other neurodegenerative disorders. Nevertheless, some limitations should be noted. Our analysis was restricted to ADNI participants, which may not fully represent broader population diversity. In addition, although our statistical associations are biologically plausible, causal mechanisms cannot be inferred without experimental validation. Future research should extend these models to longitudinal data, incorporate additional omics layers (e.g., proteomics, epigenetics), and validate findings across independent cohorts.

\section*{Acknowledgments}
Data collection and sharing for this project was funded by the Alzheimer's Disease Neuroimaging Initiative (ADNI) (National Institutes of Health Grant U01 AG024904) and DOD ADNI (Department of Defense award number W81XWH-12-2-0012). ADNI is funded by the National Institute on Aging, the National Institute of Biomedical Imaging and Bioengineering, and through generous contributions from the following: AbbVie, Alzheimer's Association; Alzheimer's Drug Discovery Foundation; Araclon Biotech; BioClinica, Inc.; Biogen; Bristol-Myers Squibb Company; CereSpir, Inc.; Cogstate; Eisai Inc.; Elan Pharmaceuticals, Inc.; Eli Lilly and Company; EuroImmun; F. Hoffmann-La Roche Ltd and its affiliated company Genentech, Inc.; Fujirebio; GE
Healthcare; IXICO Ltd.; Janssen Alzheimer's Immunotherapy Research \& Development, LLC.; Johnson \& Johnson Pharmaceutical Research \& Development LLC.; Lumosity; Lundbeck; Merck \& Co., Inc.; Meso
Scale Diagnostics, LLC.; NeuroRx Research; Neurotrack Technologies; Novartis Pharmaceuticals Corporation; Pfizer Inc.; Piramal Imaging; Servier; Takeda Pharmaceutical Company; and Transition
Therapeutics. The Canadian Institutes of Health Research is providing funds to support ADNI clinical sites in Canada. Private sector contributions are facilitated by the Foundation for the National Institutes of Health ({\tt www.fnih.org}). The grantee organization is the Northern California Institute for Research and Education,
and the study is coordinated by the Alzheimer's Therapeutic Research Institute at the University of Southern California. ADNI data are disseminated by the Laboratory for Neuro Imaging at the University of Southern California.

\bibliographystyle{unsrt}
\bibliography{biblio}

\newpage
\begin{appendix}
\section{Appendix section}

\begin{table}[htbp]
    \centering
    \caption{Patient Characteristics by Cognitive Status}
    \renewcommand{\arraystretch}{1.25} 
    \Large 
    \resizebox{\textwidth}{!}{%
    \begin{tabular}{lcccc}
    \toprule
    \textbf{Characteristic} & \textbf{Cognitively Normal (417)} & \textbf{Early Mild CI (310)} & \textbf{Late Mild CI (562)} & \textbf{Diagnosed AD (342)} \\
    \midrule
    \multicolumn{5}{l}{\textbf{Demographics}} \\
    Sex of Patient, n (\%) & & & & \\
    \quad Male & 209 (50.12\%) & 171 (55.16\%) & 344 (61.21\%) & 189 (55.26\%) \\
    \quad Female & 208 (49.88\%) & 139 (44.84\%) & 218 (38.79\%) & 153 (44.74\%) \\
    Race of Patient, n (\%) & & & & \\
    \quad White & 376 (90.17\%) & 286 (92.26\%) & 526 (93.59\%) & 317 (92.69\%) \\
    \quad Black & 30 (7.19\%) & 8 (2.58\%) & 22 (3.91\%) & 14 (4.09\%) \\
    \quad Other & 11 (2.64\%) & 16 (5.16\%) & 14 (2.42\%) & 11 (3.22\%) \\
    Formal Years of Education, Mean (SD) & 16.28 (2.73) & 15.96 (2.66) & 15.88 (2.94) & 15.18 (2.99) \\
    \midrule
    \multicolumn{5}{l}{\textbf{Risk Factors}} \\
    Baseline Age (years), Mean (SD) & 74.76 (5.73) & 71.19 (7.50) & 73.99 (7.50) & 75.03 (7.79) \\
    (Apolipoprotein E4) Genetic Status, n (\%) & & & & \\
    \quad Absent & 301 (72.53\%) & 175 (57.19\%) & 256 (45.71\%) & 113 (33.43\%) \\
    \quad Present & 114 (27.47\%) & 131 (42.81\%) & 304 (54.29\%) & 225 (66.57\%) \\
    \midrule
    \multicolumn{5}{l}{\textbf{Cognitive Exams}} \\
    ADAS (13), Mean (SD) & 9.34 (4.32) & 12.65 (5.42) & 18.66 (6.52) & 29.87 (8.05) \\
    CDRSB, Mean (SD) & 0.03 (0.13) & 1.29 (0.76) & 1.65 (0.92) & 4.39 (1.67) \\
    MMSE, Mean (SD) & 29.07 (1.12) & 28.34 (1.56) & 27.18 (1.80) & 23.22 (2.07) \\
    Rey Auditory General Test, Mean (SD) & 44.35 (9.84) & 39.55 (10.71) & 31.32 (9.51) & 22.82 (7.55) \\
    \midrule
    \multicolumn{5}{l}{\textbf{Cerebrospinal Fluid (CSF) Measurements}} \\
    Amyloid-Beta levels, Mean (SD) & 1,327.68 (660.95) & 1,178.19 (587.49) & 889.14 (490.26) & 691.44 (416.43) \\
    Tau levels, Mean (SD) & 238.45 (88.96) & 256.41 (121.74) & 308.95 (128.96) & 367.84 (144.92) \\
    Phosphorylated Tau levels, Mean (SD) & 22.00 (9.08) & 24.25 (13.69) & 30.46 (14.66) & 36.66 (15.76) \\
    \bottomrule
    \end{tabular}%
    }
    \label{tab:Biomarkers_Tab}
\end{table}

\begin{figure}[H]
    \centering
    \includegraphics[width = .9\linewidth, height = 7.85cm]{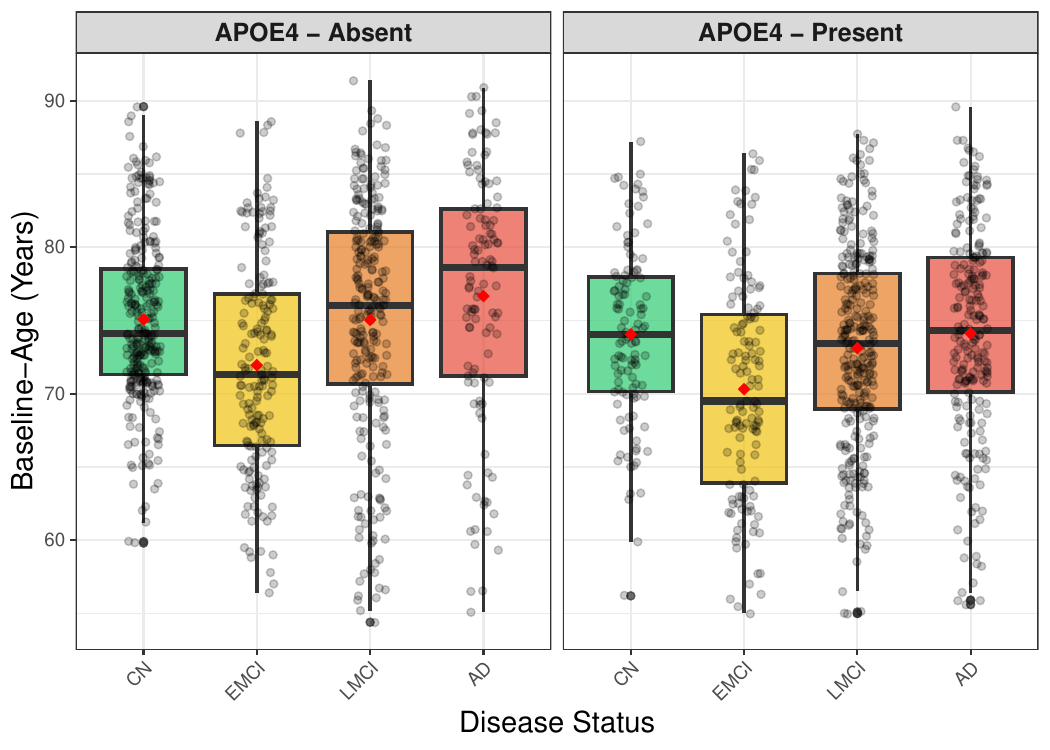}
    \caption{Distribution of age across clinical diagnostic groups (CN, EMCI, LMCI and AD) stratified by APOE4 carrier status. Boxplots display the interquartile range and median age, with individual data points shown as jittered dots for transparency of the distribution. Red diamonds indicate mean age within each group. The plot illustrates both the variation in age among diagnostic categories and potential differences by APOE4 genotype.}
    \label{fig:APOE4_Jitter}
\end{figure}

\end{appendix}

\end{document}